\documentclass[preprints,article,accept,moreauthors,pdftex]{Definitions/mdpi} 
\usepackage{amsfonts}
\usepackage{amssymb}
\usepackage{physics}
\usepackage{mathtools}
\usepackage{textcomp}
\usepackage{siunitx}
\usepackage{empheq}
\usepackage{bbold}
\usepackage{dcolumn}
\usepackage{bm}
\usepackage{wrapfig}
\usepackage{lipsum}

\firstpage{1} 
\pubvolume{1}
\issuenum{1}
\articlenumber{0}
\pubyear{2021}
\copyrightyear{2021}
\history{
}

\pdfoutput=1

\Title{Analogue non-causal Null Curves and Chronology protection in a dc-SQUID Array}

\Author{ Carlos Sab\'in $^{1,}$*\orcidA{}}
\AuthorNames{Carlos Sabín}

\address{%
$^{1}$ \quad Departamento de F\'isica Te\'orica, Universidad Aut\'onoma de Madrid, 28049 Madrid, Spain; carlos.sabin@uam.es}
\corres{Correspondence: carlos.sabin@uam.es}

\abstract{We propose an analogue quantum simulator of a 1+1 D  spacetime containing non-causal curves,in particular, null geodesics going back in time, by means of a dc-SQUID array embedded on an open superconducting transmission line. This is achieved by mimicking the spatial dependence of the metric with the propagation speed of the electromagnetic field in the simulator, which can be modulated by an external magnetic flux. We show that it is possible to simulate a spacetime region containing non-causal null geodesics, but not a full spacetime containing a chronological horizon separating regions with non-causal null geodesics and regions without them. This is in agreement with a recent suggestion of analogue-gravity chronology protection mechanism by Barcel\'o et al.}

\keyword{quantum simulation; superconducting circuits; quantum field theory in curved spacetime}

\begin{document}



\section{Introduction}

Why should we care about time travel? In 1988 , Morris, Thorne and Yurtsever \cite{PhysRevLett.61.1446} wrote:
``Normally theoretical physicists ask, "What are the laws of physics?" and/or, "What do those laws predict about the Universe?" In this Letter we ask, instead, "What constraints do the laws of physics place on the activities of an arbitrarily advanced civilization?" This will lead to some intriguing queries about the laws themselves.'' And in 1992, Thorne \cite{thome1992closed}: "Of all thought experiments, perhaps the most helpful are those that push the laws of physics in the most extreme ways". In other words, time travel seems impossible, but is this impossibility of a technological or a fundamental nature?

In principle, the existence of Closed Timelike Curves (CTCs) is allowed by General Relativity -as illustrated for instance in \cite{PhysRevLett.61.1446}- but a Chronology Protection Conjecture was proposed by Hawking \cite{chronology_protection_PhysRevD.46.603}, according to which quantum effects could prevent the formation of CTCs, therefore preventing causal violations. The conjecture is backed up by semiclassical computations within the framework of Quantum Field Theory in Curved Spacetime \cite{birrell_davies_1982}, but only a full theory of quantum gravity -yet unknown- could definitely confirm it -or, more unlikely, refute it. 
However, experimental tests of such a theory are beyond reach even in the middle/long term, and even the predictions from the semiclassical approximation of Quantum Field Theory in Curved Spacetime -such as, for instance, Hawking radiation- are hard to check experimentally. It is in this context where the use of Analogue Gravity \cite{barcelo2011analogue} becomes interesting.

For instance, classical simulators have been proposed to mimic  curved spacetimes, such as simulations of wormholes with electromagnetic metamaterials \cite{EM_wormholes_PhysRevLett.99.183901}, water \cite{water_wormholes_PhysRevD.93.084032} or magnetic metamaterials \cite{magnetic_wormhole}, superluminal motion \cite{superluminal_motion} or the event horizon of a white hole \cite{event_horizon}. Even more interesting for the problem at hand, the field of Analogue Gravity could benefit from the growing interest in Quantum Simulation, by proposing quantum simulators, in which the interplay of quantum effects with gravitational systems could in principle be analyzed. The simulation of Hawking radiation has been successfully achieved in the laboratory with Bose-Einstein condensates \cite{steinhauer} and theoretical proposals with superconducting circuits or graphene can be found in the literature \cite{quantum_vacuum_hawking_radiation_RevModPhys.84.1,PhysRevLett.103.087004,https://doi.org/10.48550/arxiv.2207.04097}. These are quantum setups in which the analysis of quantum phenomena such as quantum entanglement is crucial. Some other proposals for quantum simulations of curved spacetimes with superconducting circuits include traversable wormholes and other exotic spacetimes containing CTCs \cite{bose_wormhole_PhysRevD.97.044045, squid_wormhole_PhysRevD.94.081501,CTCs_Mart_n_V_zquez_2020,exotic_spacetimes_Sab_n_2018}. These works already hint at the possibility of an analogue chronology protection mechanism, which prevent the formation of "time machines'' even in an analogue system. This notion has been recently formalized -mostly within Bose-Einstein condensates setups- by Barceló et al. \cite{barcelo2022chronology}. They show that spacetimes containing non-causal curves which go back in time can certainly be simulated in analogue gravity systems, but it is impossible to simulate a spacetime containing both causal and non-causal regions separated by a chronological horizon. 

In this work, we analyse the ideas of Barceló et al. within the framework developed in \cite{exotic_spacetimes_Sab_n_2018}, which proposes the simulation of 1+1 dimensional sections of exotic spacetimes using a superconducting circuit consisting of a dc-SQUID array embedded in an open microwave transmission line \cite{dynamical_casimir_effect,squid_array_Haviland2000,squid_array_Watanabe_PhysRevB.67.094505,squid_array_ergul_PhysRevB.88.104501}. We apply this toolbox to simulate a spacetime which is tailor-made to contain null geodesics going back in time.


\section{Simulator}

In a dc-Superconducting Quantum Interference Device (dc-SQUID) array \cite{simoen_2015,jjs_squids_You2011} embedded in an open transmission line, the effective propagation speed of the quantum electromagnetic field can be spatially and/or temporally modulated through the inductance per unit length of the circuit, which in turn depends on the external magnetic flux:
\begin{equation}
    c=\frac{1}{\sqrt{CL}}
\end{equation}
is the effective speed of light through the array, where $C$ and $L$ are the capacitance and inductance per unit length of the array, respectively, with \cite{simoen_2015}
\begin{equation}
\label{eq:inductance}
    L_s (\phi_{\text{ext}}) = \frac{\phi_0}{4\pi I_c \abs{\cos{\frac{\pi \phi_{\text{ext}}}{\phi_0}}} \cos \psi}
\end{equation}
the inductance of a single SQUID, where $I_c$ is its critical current, $\phi_0=h/2e$ is the magnetic flux quantum, $\phi_{\text{ext}}$ is the external magnetic flux and $\psi$ is the SQUID phase difference. In the weak-signal limit regime,  $\cos\psi\simeq 1$ \cite{squid_wormhole_PhysRevD.94.081501, CTCs_Mart_n_V_zquez_2020,exotic_spacetimes_Sab_n_2018}.

In the absence of external flux the effective speed of propagation would be
\begin{equation}
    c_0\equiv c(\phi_{\text{ext}}=0)=\frac{\epsilon}{\sqrt{C_s L_s(\phi_{\text{ext}}=0)}}=\epsilon \sqrt{\frac{4\pi I_c}{\phi_0 C_s}} \,,
\end{equation}
where $\epsilon$ is the length of the SQUID. Then,
\begin{equation}
    c(\phi_{\text{ext}})=c_0 \sqrt{\abs{\cos{\frac{\pi \phi_{\text{ext}}}{\phi_0}}}}\label{eq:speedflux}
\end{equation}
gives us the speed of light as a function of the external magnetic flux. Please notice that, by construction $c(\phi_{ext})>0$.

We can split the flux into DC and AC contributions
\begin{equation}
    \phi_{\text{ext}}(x,t) = \phi_{\text{ext}}^{\text{DC}} + \phi_{\text{ext}}^{\text{AC}}(x,t) \,.
\end{equation}
We can use the DC contribution to reduce the effective speed of light of the array background \cite{exotic_spacetimes_Sab_n_2018}, so we can define
\begin{equation}
    \tilde{c}_0^2 \equiv c^2(\phi_{\text{ext}}^{\text{AC}}=0) = c_0^2 \abs{\cos{\frac{\pi \phi_{\text{ext}}^{\text{DC}}}{\phi_0}}} \,,
\end{equation}
 and therefore,
\begin{equation}
\label{eq:total_speed_light_squid}
    c^2(\phi_{\text{ext}}) = \tilde{c}_0^2(\phi_{\text{ext}}^{\text{DC}}) \tilde{c}^2(\phi_{\text{ext}}) \,,
\end{equation}
where
\begin{equation}
\label{eq:speed_light_squid}
    \tilde{c}^2(\phi_{\text{ext}}) = \abs{\sec{\frac{\pi \phi_{\text{ext}}^{\text{DC}}}{\phi_0}}} \abs{\cos{\frac{\pi \phi_{\text{ext}}}{\phi_0}}} \,.
\end{equation}
We have to keep in mind that, when we have squared Eq. (\ref{eq:speedflux}) in order to find the flux profile, we have introduced spurious solutions corresponding to $c(\phi_{ext})=-c_0\sqrt{\abs{\cos{\frac{\pi\phi_{ext}}{\phi_0}}}}$, which are not allowed in our setup.

Finally, if $\cos{\pi \phi_{\text{ext}}/\phi_0}>0 $, we have:
\begin{equation}
    \label{eq:flux_effective_speed}
    \frac{\pi\phi_{\text{ext}}^{\text{AC}}}{\phi_0} = \arccos \left[ \cos \left( \frac{\pi\phi_{\text{ext}}^{\text{DC}}}{\phi_0} \right) \tilde{c}^2 \right] - \frac{\pi\phi_{\text{ext}}^{\text{DC}}}{\phi_0}\,,
\end{equation}
while if 
$\cos{\pi \phi_{\text{ext}}/\phi_0}<0 $, then:
\begin{equation}
    \label{eq:flux_effective_speed2}
    \frac{\pi\phi_{\text{ext}}^{\text{AC}}}{\phi_0} = \arccos \left[ -\cos \left( \frac{\pi\phi_{\text{ext}}^{\text{DC}}}{\phi_0} \right) \tilde{c}^2 \right] - \frac{\pi\phi_{\text{ext}}^{\text{DC}}}{\phi_0}\,.
\end{equation}

In this work, we will assume $\phi_{\text{ext}}^{\text{DC}}=0$ and $\phi_{\text{ext}}^{\text{AC}}(x,t)=\phi_{\text{ext}}^{\text{AC}}(x)$, so $\phi_{\text{ext}}(x,t)=\phi_{\text{ext}}(x)$. Therefore, we simply have that, if $\cos{\pi \phi_{\text{ext}}/\phi_0}>0 $
\begin{equation}
    \label{eq:flux_effective_speed3}
    \frac{\pi\phi_{\text{ext}}}{\phi_0} = \arccos   
    (\tilde{c}^2),
\end{equation}
and, for $\cos{\pi \phi_{\text{ext}}/\phi_0}<0 $, then:
\begin{equation}
    \label{eq:flux_effective_speed4}
    \frac{\pi\phi_{\text{ext}}}{\phi_0} = \arccos   
    (-\tilde{c}^2),
\end{equation}



\section{A spacetime with non-causal null geodesics}

We consider the following 1+1 D line element:
\begin{equation}
    ds^2=-c^2(x) dt^2+dx^2.\label{eq:et}
\end{equation}
As it is well-known, all 1+1 D spacetimes are flat spacetimes \cite{collas1977general}. However, we do not consider Eq. (\ref{eq:et}) as a spacetime per se, but as a section of a a 3+1 spacetime \cite{exotic_spacetimes_Sab_n_2018}. Note that all 1+1 static metrics are conformal to Eq. (\ref{eq:et}).

The null geodesics of this spacetime are obviously given by $\pm c(x)$. Now if we consider a trajectory from left to right from  $x_1$ to $x_2$ and back (then $x_2>x_1$), we have that the duration of the trajectory is
\begin{equation}
T=\int_{x_1}^{x_2}\frac{dx}{c(x)}- \int_{x_2}^{x_1}\frac{dx}{c(x)}=2 \int_{x_1}^{x_2}\frac{dx}{c(x)}.
    \label{eq:trajtime}
\end{equation}
Therefore
\begin{equation}
T= 2\, (T(x_2)-T(x_1))\label{eq:trajtime2}    
\end{equation}
where
\begin{equation}
T(x)= \int\frac{dx}{c(x)}.\label{eq:trajtime3}    
\end{equation}
Therefore, we have that if $T(x_1)>T(x_2)$, then $T<0$ and we are going "back in time''. If we choose
\begin{equation}
    c(x)=-c\frac{x^3}{2\,a^3}\label{eq:speed},
\end{equation}
where $a$ is a positive constant with spatial dimensions,then $c(x)>0$ if $x<0$, and we have 
\begin{equation}
T(x)=\frac{a^3}{c\,x^2}\label{eq:trajtime4},
\end{equation}
and therefore
\begin{equation}
    T= \frac{2\,a^3}{c}\left(\frac{1}{x_2^2}-\frac{1}{x_1^2}\right) <0.\label{eq:trajtime5}
\end{equation}

For $x>0$, Eq.(\ref{eq:speed}) becomes negative and therefore it corresponds to null geodesics travelling from left to right, therefore $T>0$ since now $x_2<x_1$. Thus, we have a spacetime region with non-causal null geodesics for $x<0$ and another one with causal curves for $x>0$, separated by the chronological horizon at $x=0$. 

The choice of $c(x)$ in Eq. (\ref{eq:speed}), which in turn is the choice of simulated spacetime, is only motivated by our need of getting $T<0$ in Eq. (\ref{eq:trajtime}) and does not have any physical meaning beyond that. While there might be other choices with the same property -for instance, higher odd powers of $x$- we have not been able to come up with any simpler example. Moreover, our conclusions below would be similar.

\section{Analogue quantum simulation of a spacetime with non-causal null geodesics}

Can we implement an analogue quantum simulation of the spacetime described in the previous section? In order to check that, we need to plug Eq. (\ref{eq:speed}) into Eqs. (\ref{eq:flux_effective_speed3}) and (\ref{eq:flux_effective_speed4}). 
Then we find that, for $\cos{\pi \phi_{\text{ext}}/\phi_0}>0 $ we have:
\begin{equation}
    \label{eq:flux_effective_speed5}
    \frac{\pi\phi_{\text{ext}}}{\phi_0} = \arccos   
    (\frac{x^6}{4\,a^6}).
\end{equation}
Then we are limited by $0 \leq \frac{x^6}{4\,a^6}\leq 1$, which entails $0 \leq x\leq a\,2^{1/3}$. However, as discussed above, we have the extra limitation that we are only allowed to consider $c(x)>0$, therefore we cannot simulate Eq. (\ref{eq:speed})  for $x>0$.

For $\cos{\pi \phi_{\text{ext}}/\phi_0}<0 $ we have:
\begin{equation}
    \label{eq:flux_effective_speed5}
    \frac{\pi\phi_{\text{ext}}}{\phi_0} = \arccos   
    (-\frac{x^6}{4\,a^6}).
\end{equation}
Then we are limited by $-1 \leq -\frac{x^6}{4\,a^6}\leq 0$, which entails $a\,(-2)^{1/3} \leq x\leq 0$. Therefore, the value of the positive constant $a$ determines the size of the region that we are able to simulate. Additionally, we have to take into account what happens when we approach the chronological horizon at $x=0$. We have that $c(0)=0$, which means that $\phi_{ext}=\phi_0/2$ -infinite inductance. At this critical value, quantum fluctuations would appear due to the very high impedance of the electromagnetic environment. If a large region of the array is close to this limit, this could lead to large fluctuations in the superconducting phase $\psi$, breaking down our approximation $\cos \psi \simeq 1$ and preventing the system to be in the superconducting phase \cite{squid_wormhole_PhysRevD.94.081501, exotic_spacetimes_Sab_n_2018,squid_array_Haviland2000}. Therefore, we should try to avoid it by keeping as few SQUIDs as possible close to this limit, ideally a single SQUID, by cleverly choosing the values of $a$. However, this would also have the effect of diminishing the size of the whole region that we are able to simulate, not only the problematic part. 
\begin{figure}
\includegraphics[width=15.5 cm]{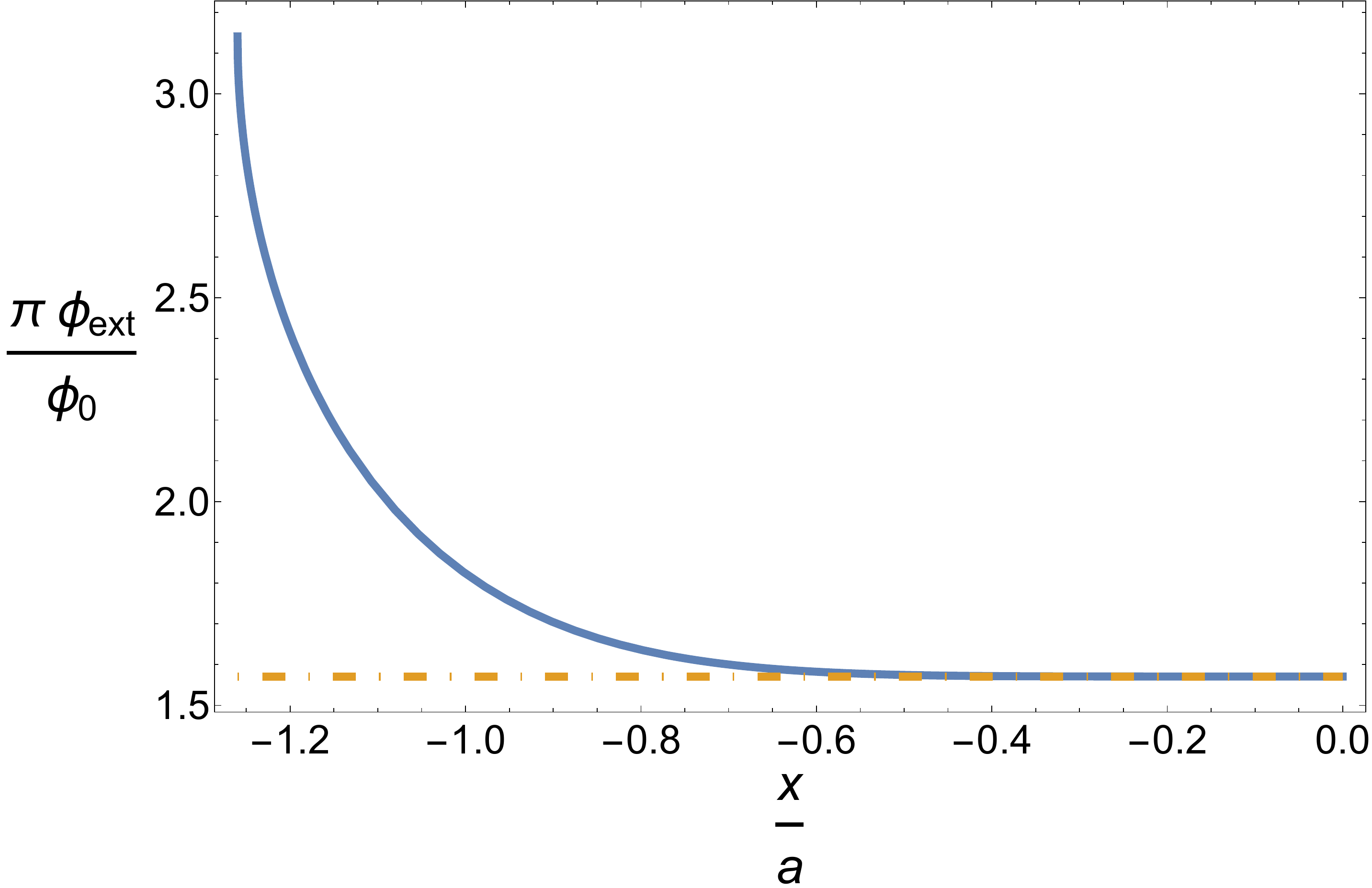}
\caption{External magnetic flux $\pi\phi_{\text{ext}}/\phi_0$ versus the adimensional distance $x/a$. The dashed line represents the threshold value $\pi\phi_{\text{ext}}/\phi_0=\pi/2$.}
\centering
\label{fig:1}
\end{figure}
In Figure \ref{fig:1}, we see that the flux is very close to the critical value in a large region of the parameter space, preventing us from getting close to the chronological horizon. We can safely simulate the spacetime in the region going from $-2^{1/3}\,a$ which corresponds to $\phi_{ext}=\phi_0$ -flat-spacetime $c$- to approximately $-0.9\,a$ , that is $\phi_{ext}\approx3/5\phi_0$.

Therefore, we see that it is possible to simulate a (small) region of spacetime containing non-causal null geodesics. However, these curves are of the type that Barceló et al. \cite{barcelo2022chronology} deemed as ``trivial'', since they appear in all the simulated spacetime: there is no "causal" region from which we could travel to the non-causal one.

\section{Summary and conclusions}

In conclusion,we have applied the procedure of  \cite{exotic_spacetimes_Sab_n_2018} to a possible quantum simulation of a 1+1 D spacetime tailor-made to contain trajectories going back in time. We have found a small region of parameters where the simulation is possible. In this region, there would be null geodesics travelling back in time. However, we are not able to simulate a full spacetime containing both this non-causal part together with a standard causal part where null geodesics go forward in time. In our setup, this would require to simulate negative field propagation speeds, which is not allowed by the physical features of our quantum simulator. Moreover, we are not even able to get close to the point where $c=0$. This point would be a chronological horizon separating the non causal part from the -in our setup impossible- causal part. We find that not only at the horizon we would have a point of infinite inductance, but that there is a large region of the simulated spacetime where the flux would be close to the critical value generating infinite inductance. This would trigger large quantum fluctuations of the superconducting phase, preventing any chance of simulating the horizon. These divergencies can be related with the ones in Barceló et al. \cite{barcelo2022chronology}, which found a similar analogue chronology protection mechanism preventing the construction of spacetimes with chronological horizons, in which we could travel from causal to non-causal regions. Our non-causal null geodesics would lie within the class that they denote as ``trivial'', since we could always blame them to the particular definition of time within a non-causal bubble, which would be impossible to leave or enter from a causal one using the same definition of time.

This example illustrates the existence of analogue chronology protection mechanisms in the quantum simulation of curved spacetimes. Recalling Hawking's famous statement \cite{chronology_protection_PhysRevD.46.603}: ``It seems there is a chronology protection agency, which prevents the appearance of closed timelike curves and so makes the universe safe for historians.'' Even for analogue historians.




\vspace{6pt} 




\funding{CS has received financial support through the Ram\'on y Cajal programme (RYC2019-028014-I).}


\conflictsofinterest{The authors declare no conflict of interest.} 




\begin{thebibliography}{-------}
\providecommand{\natexlab}[1]{#1}

\bibitem[Morris \em{et~al.}(1988)Morris, Thorne, and
  Yurtsever]{PhysRevLett.61.1446}
Morris, M.S.; Thorne, K.S.; Yurtsever, U.
\newblock Wormholes, Time Machines, and the Weak Energy Condition.
\newblock {\em Phys. Rev. Lett.} {\bf 1988}, {\em 61},~1446--1449.
\newblock
  doi:{\changeurlcolor{black}\href{https://doi.org/10.1103/PhysRevLett.61.1446}{\detokenize{10.1103/PhysRevLett.61.1446}}}.

\bibitem[Thome(1992)]{thome1992closed}
Thome, K.S.
\newblock Closed timelike curves.
\newblock {\em General Relativity and Gravitation} {\bf 1992}, p. 295.

\bibitem[Hawking(1992)]{chronology_protection_PhysRevD.46.603}
Hawking, S.W.
\newblock Chronology protection conjecture.
\newblock {\em Phys. Rev. D} {\bf 1992}, {\em 46},~603--611.
\newblock
  doi:{\changeurlcolor{black}\href{https://doi.org/10.1103/PhysRevD.46.603}{\detokenize{10.1103/PhysRevD.46.603}}}.

\bibitem[Birrell and Davies(1982)]{birrell_davies_1982}
Birrell, N.D.; Davies, P.C.W.
\newblock {\em Quantum Fields in Curved Space}; Cambridge Monographs on
  Mathematical Physics, Cambridge University Press,  1982.
\newblock
  doi:{\changeurlcolor{black}\href{https://doi.org/10.1017/CBO9780511622632}{\detokenize{10.1017/CBO9780511622632}}}.

\bibitem[Barcel{\'o} \em{et~al.}(2011)Barcel{\'o}, Liberati, and
  Visser]{barcelo2011analogue}
Barcel{\'o}, C.; Liberati, S.; Visser, M.
\newblock Analogue gravity.
\newblock {\em Living reviews in relativity} {\bf 2011}, {\em 14},~1--159.

\bibitem[Greenleaf \em{et~al.}(2007)Greenleaf, Kurylev, Lassas, and
  Uhlmann]{EM_wormholes_PhysRevLett.99.183901}
Greenleaf, A.; Kurylev, Y.; Lassas, M.; Uhlmann, G.
\newblock Electromagnetic Wormholes and Virtual Magnetic Monopoles from
  Metamaterials.
\newblock {\em Phys. Rev. Lett.} {\bf 2007}, {\em 99},~183901.
\newblock
  doi:{\changeurlcolor{black}\href{https://doi.org/10.1103/PhysRevLett.99.183901}{\detokenize{10.1103/PhysRevLett.99.183901}}}.

\bibitem[Peloquin \em{et~al.}(2016)Peloquin, Euv\'e, Philbin, and
  Rousseaux]{water_wormholes_PhysRevD.93.084032}
Peloquin, C.; Euv\'e, L.P.; Philbin, T.; Rousseaux, G.
\newblock Analog wormholes and black hole laser effects in hydrodynamics.
\newblock {\em Phys. Rev. D} {\bf 2016}, {\em 93},~084032.
\newblock
  doi:{\changeurlcolor{black}\href{https://doi.org/10.1103/PhysRevD.93.084032}{\detokenize{10.1103/PhysRevD.93.084032}}}.

\bibitem[Prat-Camps \em{et~al.}(2015)Prat-Camps, Navau, and
  Sanchez]{magnetic_wormhole}
Prat-Camps, J.; Navau, C.; Sanchez, A.
\newblock A Magnetic Wormhole.
\newblock {\em Scientific Reports} {\bf 2015}, {\em 5},~12488.
\newblock
  doi:{\changeurlcolor{black}\href{https://doi.org/10.1038/srep12488}{\detokenize{10.1038/srep12488}}}.

\bibitem[Clerici \em{et~al.}(2016)Clerici, Spalding, Warburton, Lyons,
  Aniculaesei, Richards, Leach, Henderson, and Faccio]{superluminal_motion}
Clerici, M.; Spalding, G.C.; Warburton, R.; Lyons, A.; Aniculaesei, C.;
  Richards, J.M.; Leach, J.; Henderson, R.; Faccio, D.
\newblock Observation of image pair creation and annihilation from superluminal
  scattering sources.
\newblock {\em Science Advances} {\bf 2016}, {\em 2},~e1501691.
\newblock
  doi:{\changeurlcolor{black}\href{https://doi.org/10.1126/sciadv.1501691}{\detokenize{10.1126/sciadv.1501691}}}.

\bibitem[Philbin \em{et~al.}(2008)Philbin, Kuklewicz, Robertson, Hill,
  K~König, and Leonhardt]{event_horizon}
Philbin, T.G.; Kuklewicz, C.; Robertson, S.; Hill, S.; K~König, F.; Leonhardt,
  U.
\newblock Fiber-Optical Analog of the Event Horizon.
\newblock {\em Science} {\bf 2008}, {\em 319},~1367.
\newblock
  doi:{\changeurlcolor{black}\href{https://doi.org/10.1126/science.1153625}{\detokenize{10.1126/science.1153625}}}.

\bibitem[Steinahuer(2016)]{steinhauer}
Steinahuer, J.
\newblock Observation of quantum Hawking radiation and its entanglement in an
  analogue black hole.
\newblock {\em Nature Phys} {\bf 2016}, {\em 12},~959.
\newblock
  doi:{\changeurlcolor{black}\href{https://doi.org/10.1038/nphys3863}{\detokenize{10.1038/nphys3863}}}.

\bibitem[Nation \em{et~al.}(2012)Nation, Johansson, Blencowe, and
  Nori]{quantum_vacuum_hawking_radiation_RevModPhys.84.1}
Nation, P.D.; Johansson, J.R.; Blencowe, M.P.; Nori, F.
\newblock Colloquium: Stimulating uncertainty: Amplifying the quantum vacuum
  with superconducting circuits.
\newblock {\em Rev. Mod. Phys.} {\bf 2012}, {\em 84},~1--24.
\newblock
  doi:{\changeurlcolor{black}\href{https://doi.org/10.1103/RevModPhys.84.1}{\detokenize{10.1103/RevModPhys.84.1}}}.

\bibitem[Nation \em{et~al.}(2009)Nation, Blencowe, Rimberg, and
  Buks]{PhysRevLett.103.087004}
Nation, P.D.; Blencowe, M.P.; Rimberg, A.J.; Buks, E.
\newblock Analogue Hawking Radiation in a dc-SQUID Array Transmission Line.
\newblock {\em Phys. Rev. Lett.} {\bf 2009}, {\em 103},~087004.
\newblock
  doi:{\changeurlcolor{black}\href{https://doi.org/10.1103/PhysRevLett.103.087004}{\detokenize{10.1103/PhysRevLett.103.087004}}}.

\bibitem[Acquaviva \em{et~al.}(2022)Acquaviva, Iorio, Pais, and
  Smaldone]{https://doi.org/10.48550/arxiv.2207.04097}
Acquaviva, G.; Iorio, A.; Pais, P.; Smaldone, L.
\newblock Hunting Quantum Gravity with Analogs: the case of graphene,  2022.
\newblock
  doi:{\changeurlcolor{black}\href{https://doi.org/10.48550/ARXIV.2207.04097}{\detokenize{10.48550/ARXIV.2207.04097}}}.

\bibitem[Mateos and Sab\'{\i}n(2018)]{bose_wormhole_PhysRevD.97.044045}
Mateos, J.; Sab\'{\i}n, C.
\newblock Quantum simulation of traversable wormhole spacetimes in a
  Bose-Einstein condensate.
\newblock {\em Phys. Rev. D} {\bf 2018}, {\em 97},~044045.
\newblock
  doi:{\changeurlcolor{black}\href{https://doi.org/10.1103/PhysRevD.97.044045}{\detokenize{10.1103/PhysRevD.97.044045}}}.

\bibitem[Sab\'{\i}n(2016)]{squid_wormhole_PhysRevD.94.081501}
Sab\'{\i}n, C.
\newblock Quantum simulation of traversable wormhole spacetimes in a dc-SQUID
  array.
\newblock {\em Phys. Rev. D} {\bf 2016}, {\em 94},~081501.
\newblock
  doi:{\changeurlcolor{black}\href{https://doi.org/10.1103/PhysRevD.94.081501}{\detokenize{10.1103/PhysRevD.94.081501}}}.

\bibitem[Mart{\'{\i}}n-V{\'{a}}zquez and
  Sab{\'{\i}}n(2020)]{CTCs_Mart_n_V_zquez_2020}
Mart{\'{\i}}n-V{\'{a}}zquez, G.; Sab{\'{\i}}n, C.
\newblock Closed timelike curves and chronology protection in quantum and
  classical simulators.
\newblock {\em Classical and Quantum Gravity} {\bf 2020}, {\em 37},~045013.
\newblock
  doi:{\changeurlcolor{black}\href{https://doi.org/10.1088/1361-6382/ab5f3f}{\detokenize{10.1088/1361-6382/ab5f3f}}}.

\bibitem[Sab{\'{\i}}n(2018)]{exotic_spacetimes_Sab_n_2018}
Sab{\'{\i}}n, C.
\newblock One-dimensional sections of exotic spacetimes with superconducting
  circuits.
\newblock {\em New Journal of Physics} {\bf 2018}, {\em 20},~053028.
\newblock
  doi:{\changeurlcolor{black}\href{https://doi.org/10.1088/1367-2630/aac0db}{\detokenize{10.1088/1367-2630/aac0db}}}.

\bibitem[Barcel{\'o} \em{et~al.}(2022)Barcel{\'o}, Eguia~S{\'a}nchez,
  Garc{\'\i}a-Moreno, and Jannes]{barcelo2022chronology}
Barcel{\'o}, C.; Eguia~S{\'a}nchez, J.; Garc{\'\i}a-Moreno, G.; Jannes, G.
\newblock Chronology protection implementation in analogue gravity.
\newblock {\em The European Physical Journal C} {\bf 2022}, {\em 82},~1--17.

\bibitem[L{\"a}hteenm{\"a}ki \em{et~al.}(2013)L{\"a}hteenm{\"a}ki, Paraoanu,
  Hassel, and Hakonen]{dynamical_casimir_effect}
L{\"a}hteenm{\"a}ki, P.; Paraoanu, G.S.; Hassel, J.; Hakonen, P.J.
\newblock Dynamical Casimir effect in a Josephson metamaterial.
\newblock {\em Proceedings of the National Academy of Sciences} {\bf 2013},
  {\em 110},~4234--4238.
\newblock
  doi:{\changeurlcolor{black}\href{https://doi.org/10.1073/pnas.1212705110}{\detokenize{10.1073/pnas.1212705110}}}.

\bibitem[Haviland \em{et~al.}(2000)Haviland, Andersson, and
  {\AA}gren]{squid_array_Haviland2000}
Haviland, D.B.; Andersson, K.; {\AA}gren, P.
\newblock Superconducting and Insulating Behavior in One-Dimensional Josephson
  Junction Arrays.
\newblock {\em Journal of Low Temperature Physics} {\bf 2000}, {\em
  118},~733--749.
\newblock
  doi:{\changeurlcolor{black}\href{https://doi.org/10.1023/A:1004603814529}{\detokenize{10.1023/A:1004603814529}}}.

\bibitem[Watanabe and Haviland(2003)]{squid_array_Watanabe_PhysRevB.67.094505}
Watanabe, M.; Haviland, D.B.
\newblock Quantum effects in small-capacitance single Josephson junctions.
\newblock {\em Phys. Rev. B} {\bf 2003}, {\em 67},~094505.
\newblock
  doi:{\changeurlcolor{black}\href{https://doi.org/10.1103/PhysRevB.67.094505}{\detokenize{10.1103/PhysRevB.67.094505}}}.

\bibitem[Erg\"ul \em{et~al.}(2013)Erg\"ul, Schaeffer, Lindblom, Haviland,
  Lidmar, and Johansson]{squid_array_ergul_PhysRevB.88.104501}
Erg\"ul, A.; Schaeffer, D.; Lindblom, M.; Haviland, D.B.; Lidmar, J.;
  Johansson, J.
\newblock Phase sticking in one-dimensional Josephson junction chains.
\newblock {\em Phys. Rev. B} {\bf 2013}, {\em 88},~104501.
\newblock
  doi:{\changeurlcolor{black}\href{https://doi.org/10.1103/PhysRevB.88.104501}{\detokenize{10.1103/PhysRevB.88.104501}}}.

\bibitem[Simoen(2015)]{simoen_2015}
Simoen, M.
\newblock Parametric Interactions with Signals and the Vacuum.
\newblock PhD thesis, Chalmers University of Technology,  2015.

\bibitem[You and Nori(2011)]{jjs_squids_You2011}
You, J.Q.; Nori, F.
\newblock Atomic physics and quantum optics using superconducting circuits.
\newblock {\em Nature} {\bf 2011}, {\em 474},~589--597.
\newblock
  doi:{\changeurlcolor{black}\href{https://doi.org/10.1038/nature10122}{\detokenize{10.1038/nature10122}}}.

\bibitem[Collas(1977)]{collas1977general}
Collas, P.
\newblock General relativity in two-and three-dimensional space--times.
\newblock {\em American Journal of Physics} {\bf 1977}, {\em 45},~833--837.

\end{thebibliography}
\end{document}